\documentclass[reqno]{amsart}%

\usepackage{amsmath}
\usepackage{amsfonts,amssymb}
\usepackage{cases,graphics}
\usepackage{verbatim}

\newtheorem{theorem}{Theorem}
\newtheorem{lemma}{Lemma}
\newtheorem{proposition}{Proposition}
\newtheorem{corollary}{Corollary}

\theoremstyle{definition} 
\newtheorem{definition}{Definition}
\theoremstyle{remark} 
\newtheorem{remark}{Remark}

\newcommand{\mynoindent}{} 
\newcommand{\enproof}{\hfill $\Box$\vspace*{1ex}}
\newcommand{\enproofwosp}{\hfill $\Box$} 

\newcommand{\mymathbb}[1]{{\mathbb #1}} 

\newcommand{\SRN}{\mymathbb{R}}
\newcommand{\SCN}{\mymathbb{C}}
\newcommand{\SNN}{\mymathbb{N}}
\newcommand{\SINT}{\mymathbb{Z}}

\newcommand{\imu}{i} 
\newcommand{\nap}{e} 

\newcommand{\transp}{^{\rm T}}
\newcommand{\abs}[1]{|#1|}

\newcommand{\mfrac}[2]{\mbox{$\frac{#1}{#2}$}}
\newcommand{\sfrac}[2]{\mfrac{#1}{#2}} 
\newcommand{\slfrac}[2]{#1/#2}
\newcommand{\morfrac}[2]{\frac{#1}{#2}} 

\newcommand{\numK}{k}
\newcommand{\varnu}{j} 
\newcommand{\myv}[1]{\hat{#1}} 
\newcommand{\myvec}[1]{#1} 

\newcommand{\thDR}{\hat{R}}

\newcommand{\anga}{\alpha} 
\newcommand{\angb}{\beta} 
\newcommand{\angc}{\gamma} 
\newcommand{\angg}{\phi} 
\newcommand{\cR}{\mathcal{R}} 
\newcommand{\cD}{\mathcal{A}} 

\newcommand{\rinnpr}[2]{#1\transp #2} 
\newcommand{\rinnprsp}[2]{#1\hspace{0.15ex} \transp #2} 
\newcommand{\vargv}{v} 
\newcommand{\varvn}{n} 
\newcommand{\varM}[2]{M_{#1}^{#2}}
\newcommand{\varMcomb}[1]{M_{#1}}
\newcommand{\myk}{k}
\newcommand{\vark}{k} 

\newcommand{\myb}{b} 
\newcommand{\varu}{u} 
\newcommand{\genD}{A} 

\newcommand{\mydisplaystyle}{}

\newcommand{\refapp}[1]{Appendix~\ref{#1}}
\newcommand{\refappcomma}[1]{Appendix~\ref{#1}}
\newcommand{\refapps}[2]{Appendices~\ref{#1} and \ref{#2}}

\newcommand{\sectionapp}[1]{\section{#1}} 

\newcommand{\refroyalsub}[2]{\ref{#2}}
\newcommand{\refappparen}[1]{(Appendix~\ref{#1})}

\newcommand{\mtilde}[1]{#1}

\newcommand{\varkapp}{k}

\begin{document}

\title[The Minimum Number of Rotations About Two Axes]{The Minimum Number of Rotations About Two Axes for Constructing an Arbitrarily Fixed Rotation}

\author{%
Mitsuru Hamada} 
\thanks{%
The author is with 
Tamagawa University,
Tamagawa-gakuen 1-chome, Machida, Tokyo 194-8610, Japan.}


\date{}

\dedicatory{}

\begin{abstract}
For any pair of three-dimensional real unit vectors $\myv{m}$ and $\myv{\varvn}$ with $\abs{\rinnpr{\myv{m}}{\myv{\varvn}}} < 1$ and
any rotation $U$, let $N_{\myv{m},\myv{n}}(U)$ denote 
the least value of a positive integer $\varkapp$ such that $U$
can be decomposed into 
a product of $\varkapp$ rotations about either $\myv{m}$ or $\myv{\varvn}$.
This work gives the number $N_{\myv{m},\myv{n}}(U)$
as a function of $U$. 
Here a rotation means an element $D$ of the special orthogonal group ${\rm SO}(3)$
or an element of the special unitary group ${\rm SU}(2)$ that corresponds to $D$.
Decompositions 
of $U$ attaining the minimum number $N_{\myv{m},\myv{n}}(U)$
are also given explicitly.
\end{abstract}

\maketitle


\section{Introduction \label{ss:intro}}

In this work,
an issue on optimal constructions of rotations in the 
Euclidean space $\SRN^3$, under some restriction,
is addressed and solved.
By a rotation or rotation matrix, we usually mean an element of the special orthogonal group ${\rm SO}(3)$.
However, we follow the custom, in quantum physics, to call not only an element of ${\rm SO}(3)$ but also that of the special unitary group ${\rm SU}(2)$ 
a rotation.
This is justified by the well-known homomorphism from ${\rm SU}(2)$
onto ${\rm SO}(3)$ (Section~\refroyalsub{ss:prel}{sssub:homomorphism}). 
Given a pair of three-dimensional real unit vectors $\myv{m}$ and $\myv{\varvn}$ with $\abs{\rinnpr{\myv{m}}{\myv{\varvn}}} < 1$, 
where $\myv{m}\transp$ denotes the transpose of $\myv{m}$,
let $N_{\myv{m},\myv{n}}(\cD)$ denote 
the least value of a positive integer $k$ such that any
rotation in $\cD$ can be decomposed into (constructed as) a product of $k$ rotations about either $\myv{m}$ or $\myv{\varvn}$,
where $\cD = {\rm SU}(2),{\rm SO}(3)$.
It is known that 
$
N_{\myv{m},\myv{n}}\big({\rm SO}(3)\big) = N_{\myv{m},\myv{n}}\big({\rm SU}(2)\big) = 
\lceil  \pi / \arccos \abs{\rinnpr{\myv{m}}{\myv{\varvn}}} \rceil +1
$ 
for any pair of three-dimensional real unit vectors $\myv{m}$ and $\myv{\varvn}$ with $\abs{\rinnpr{\myv{m}}{\myv{\varvn}}} < 1$~\cite{Lowenthal71,Lowenthal72}.

Then, a natural question arises:
What is the least value, $N_{\myv{m},\myv{n}}(U)$, of a positive integer $\varkapp$ such that an arbitrarily fixed rotation $U$
can be decomposed into 
a product of $\varkapp$ rotations about either $\myv{m}$ or $\myv{\varvn}$? 
In this work,
the minimum number $N_{\myv{m},\myv{n}}(U)$ is given as an explicit 
function of $U$, where $U$ is expressed in terms of 
parameters
known as Euler angles~\cite{Wigner,BiedenharnLouck}. 
Moreover,
optimal, i.e., minimum-achieving decompositions (constructions) of any fixed element $U \in {\rm SU}(2)$
are presented explicitly.

In this work, not only explicit constructions 
but also
simple inequalities on geometric quantities, which directly show lower bounds on the number of constituent rotations, will be presented.
Remarkably,
the proposed explicit constructions meet the obtained lower bounds, which shows both the optimality of the constructions and the tightness of the bounds.

The results in this work were obtained before the author came to know
Lowenthal's formula on 
$N_{\myv{m},\myv{n}}\big({\rm SO}(3)\big)$~\cite{Lowenthal71,Lowenthal72} and
a related result~\cite{DA04}. Prior to the present work, the work~\cite{DA04} has treated the issue of
determining $N_{\myv{m},\myv{n}}(D)$, $D \in {\rm SO}(3)$. 
The interesting result~\cite{DA04}, however, gave
$N_{\myv{m},\myv{n}}(D)$, $D \in {\rm SO}(3)$, 
only algorithmically (with the largest index of a sequence of real numbers 
with some property). 
The distinctive features of the present work include the following:
$N_{\myv{m},\myv{n}}(U)$ is given in terms of an explicit function of parameters of $U\in{\rm SU}(2)$;
explicit optimal decompositions are presented; 
this work's results on $N_{\myv{m},\myv{n}}(U)$ imply Lowenthal's formula on $N_{\myv{m},\myv{n}}\big({\rm SO}(3)\big)$ in a consistent self-contained manner.%
\footnote{Here the crux of the difficulty in obtaining this work's results will be explained.
Finding the minimum {\em odd}\/ number of factors needed for decomposing $U$, which is expressed with a standard parameter $\angb$ of $U$, together with minimum-achieving decompositions, was relatively easy.
The crux 
lay in obtaining a solution to attain the minimum {\em even}\/ number of factors, which was found to be expressed with a new parameter $\angb'$ eventually.}

Regarding another direction of related research, we remark that $N_{\myv{m},\myv{n}}(\cD)$ is known as the order of (uniform) generation of the Lie group $\cD$, and this notion has been extended to other Lie groups.
The interested reader is referred to relatively extensive treatments on uniform 
generation~\cite{KLowenthal75,Leite91},
where one would find that
even determining the order $N_{\myv{m},\myv{n}}\big({\rm SO}(3)\big)$ needs a special proof~\cite{Lowenthal71,Lowenthal72}, \cite[Appendix]{Leite91}.

Detailed elementary arguments below would help us dispel some confusions related to $N_{\myv{m},\myv{n}}\big({\rm SU}(2)\big)$
often found in textbooks on quantum computation. 
There, not to mention the ignorance of the fact
$N_{\myv{m},\myv{n}}\big({\rm SU}(2)\big)=\lceil  \pi / \arccos \abs{\rinnpr{\myv{m}}{\myv{\varvn}}} \rceil +1$, a wrong statement equivalent to saying that
$N_{\myv{m},\myv{n}}\big({\rm SU}(2)\big)$ were {\em three}, regardless of the choice of non-parallel vectors $\myv{m}$ and $\myv{n}$, is observed.

Regarding physics,
this work has been affected by the issue of constructing an arbitrary unitary operator on 
a 
Hilbert space
discussed in quantum physics~\cite{ReckZBB94}. This is relevant to universal gates for quantum computation~\cite{BoykinMPRV99}.
In this context, 
requiring the availability of rotations about a pair of exactly orthogonal axes seems too idealistic.
For example, consider a Hamiltonian $H$ of a quantum system represented by $\SCN^2$, and note that $H$ determines the axis of the rotations
$[c(t)]^{-1} \exp (-{\imu}t H ) \in {\rm SU}(2)$, $t\in \SRN$, where $c(t)$ is a square root of $\det \exp (-{\imu}t H)$. [Often, although not always, differences of unitary matrices (evolutions) up to scalar multiples are ignorable.]
Thus, explicit decompositions attaining the minimum $N_{\myv{m},\myv{n}}(U)$ of an arbitrary rotation $U$ for the generic vectors $\myv{m}$ and $\myv{n}$ will be useful.
For applications to control, the reader is referred to \cite{DA04} and references therein.

This paper is organised as follows.
After giving preliminaries in Section~\ref{ss:prel}, the main theorem establishing $N_{\myv{m},\myv{n}}(U)$ and explicit 
constructions of rotations are presented in Section~\ref{ss:const}. 
Then, inequalities that show limits on constructions are presented in Section~\ref{ss:limits}.
The proofs of the results of this work are presented in Section~\ref{ss:Proofs}.
Section~\ref{ss:conc} contains the conclusion.
Several arguments are relegated to appendices.

\section{Preliminaries and a Known Result \label{ss:prel}}

\subsection{Definitions}

The notation to be used includes the following: 
$\SNN$ denotes the set of strictly positive integers;
$S^2= \{ \myv{\vargv} \in \SRN^3 \mid \| \myv{\vargv} \| =1 \}$ where
$\| \myv{\vargv} \| = \sqrt{\vargv_x^2+\vargv_y^2+\vargv_z^2}$ for $\myv{\vargv}=(\vargv_x,\vargv_y,\vargv_z)\transp$;
$\lceil x \rceil$ denotes
the smallest integer not less than $x \in \SRN$.
As usual, $\arccos x \in [0,\pi]$ and $\arcsin x \in [-\pi/2,\pi/2]$ for $x\in [-1,1]$.
The Hermitian conjugate of a matrix $U$ is denoted by $U^{\dagger}$.

Throughout, $I$ denotes the $2\times 2$ identity matrix;
$X,Y$, and $Z$ denote the following Pauli matrices:
\[
X=\begin{pmatrix}
0 & 1 \\
1 & 0
\end{pmatrix},
\quad
Y=\begin{pmatrix}
0 & -\imu \\
\imu & 0
\end{pmatrix},
\quad
Z=\begin{pmatrix}
1 & 0 \\
0 & -1
\end{pmatrix} .
\]
We shall work with a matrix
\begin{equation}
R_{\myv{\vargv}}(\theta) := (\cos\mfrac{\theta}{2}) I - \imu 
(\sin\mfrac{\theta}{2})(\vargv_x X + \vargv_y Y + \vargv_z Z) \label{eq:Rtheta}
\end{equation}
where
$\myv{\vargv}=(\vargv_x,\vargv_y,\vargv_z)\transp \in S^2$
and $\theta\in\SRN$.
This represents the rotation about $\myv{\vargv}$ by angle $ \theta$ (through the homomorphism in Section~\ref{sssub:homomorphism}). 
In particular, for $\myv{y}=(0,1,0)\transp$ and $\myv{z}=(0,0,1)\transp$, we put
\[
R_{y}(\theta):=R_{\myv{y}}(\theta) 
=
\begin{pmatrix}
\cos\frac{\theta}{2} \ & - \sin\frac{\theta}{2} \\ 
\mbox{}\ \sin\frac{\theta}{2}  &  \cos\frac{\theta}{2} 
\end{pmatrix}
\quad \mbox{and}\quad R_{z}(\theta):=R_{\myv{z}}(\theta)
=
\begin{pmatrix}
\nap^{-\imu\frac{\theta}{2}} & 0 \\
0 & \nap^{\imu\frac{\theta}{2}}
\end{pmatrix} .
\]

For $\myv{m},\myv{\varvn}\in S^2$ with $\abs{\rinnpr{\myv{m}}{\myv{\varvn}}} < 1$, we define the following:
\begin{equation}
N_{\myv{m},\myv{n}}(U) := 
\min \{  \varnu \in \SNN \mid \exists V_1, 
\dots,V_{\varnu}\in \cR_{\myv{m}} \cup \cR_{\myv{\varvn}},\ U= V_1 
\cdots V_{\varnu} \} \label{eq:Nmn}
\end{equation}
for $U\in {\rm SU}(2)$, where
$ 
\cR_{\myv{\vargv}} := \{ R_{\myv{\vargv}}(\theta) \mid \theta \in \SRN \}
$, 
and
\begin{equation}
N_{\myv{m},\myv{n}} := N_{\myv{m},\myv{n}}\big({\rm SU}(2)\big) := \min \{ \myk \in \SNN \mid \forall U \in {\rm SU}(2), \ N_{\myv{m},\myv{n}}(U) \le \myk \}  .
\end{equation}

Using the homomorphism $F$ from ${\rm SU}(2)$ onto ${\rm SO}(3)$ to be defined in Section~\ref{sssub:homomorphism}, 
we put
$ 
\hat{\cR}_{\myv{\vargv}} := \big\{ F\big(R_{\myv{\vargv}}(\theta)\big) \mid \theta \in \SRN \big\}$.
We extend the definition of $N_{\myv{m},\myv{n}}$ to ${\rm SO}(3)$:
\begin{equation}
N_{\myv{m},\myv{n}}(D) := 
\min \{  \varnu \in \SNN \mid \exists \genD_1,
\dots,\genD_{\varnu}\in \hat{\cR}_{\myv{m}} \cup \hat{\cR}_{\myv{\varvn}},\ D= \genD_1 
\cdots \genD_{\varnu} \} \label{eq:NmnSO3}
\end{equation}
for $D\in {\rm SO}(3)$
and
\begin{equation}
N_{\myv{m},\myv{n}}\big({\rm SO}(3)\big) := \min \{ \myk \in \SNN \mid \forall D \in {\rm SO}(3), \ N_{\myv{m},\myv{n}}(D) \le \myk \}  .
\end{equation}

\subsection{The Maximum of the Minimum Number of Constituent Rotations Over All Target Rotations}

This work's results lead to an elementary self-contained proof of the following known theorem (Appendix~\ref{ss:proofthDecomp}).
\begin{theorem}[{\rm Lowenthal~\cite{Lowenthal71,Lowenthal72}}] 
\label{th:Decomp} 
For any $\myv{m},\myv{\varvn}\in S^2$ with
$\abs{\rinnpr{\myv{m}}{\myv{\varvn}}} < 1$,
\[
N_{\myv{m},\myv{n}}\big({\rm SO}(3)\big) = N_{\myv{m},\myv{n}}\big({\rm SU}(2)\big) = 
\Big\lceil \frac{ \pi }{\arccos \abs{\rinnpr{\myv{m}}{\myv{\varvn}}} } \Big\rceil +1.
\]
\end{theorem}


\subsection{Parameterisations of the Elements in SU(2)\label{sssub:para}}
\mbox{}\ 
The following lemma presents a well-known parameterisation of ${\rm SU}(2)$ elements.
\begin{lemma}\label{lem:para}
For any element 
$U\in{\rm SU}(2)$,
there exist some $\anga,\angc \in \SRN$,
and $\angb\in [0,\pi]$ such that
\begin{equation} 
U=
\begin{pmatrix}
\nap^{-\imu \frac{\angc+\anga}{2}}\cos\frac{\angb}{2} & 
\mbox{}\, - \nap^{\imu \frac{\angc-\anga}{2}} \sin\frac{\angb}{2}\\
\nap^{- \imu \frac{\angc-\anga}{2}} \sin\frac{\angb}{2} &   
\nap^{\imu \frac{\angc+\anga}{2}} \cos\frac{\angb}{2}
\end{pmatrix} = R_{z}(\anga) R_{y}(\angb) R_{z}(\angc) . \label{eq:uni1}
\end{equation}
\end{lemma}

The parameters $\anga,\angb$, and $\angc$ in this lemma are often called Euler angles.%
\footnote{The restriction of $\angb$ to $[0,\pi]$ does not seem common. However,
in a straightforward proof of this lemma, $\angb\in [0,\pi]$ can be chosen so that
$\cos(\angb/2)=\abs{a}$ and $\sin(\angb/2)=\abs{b}$ when the first row of $U$ is $(a,b)$. 
Also any $R_{z}(\anga') R_{y}(\angb') R_{z}(\angc')$ without this restriction
can be written as $R_{z}(\anga) R_{y}(\angb) R_{z}(\angc)$ with some $\angb\in [0,\pi]$ and $\anga,\angc\in\SRN$.
This readily follows from 
equations 
$R_{\myv{v}}(\theta+2\pi)=-R_{\myv{v}}(\theta)$, $\myv{v}\in S^2$, $\theta\in\SRN$,
and $R_z(-\pi)R_y(\angb')R_{z}(\pi)=R_y(-\angb')$, $\angb'\in\SRN$.
\label{fn:pipi}}
The lemma can be rephrased as follows:
Any matrix in ${\rm SU}(2)$ can be
written as
\begin{equation}
\begin{pmatrix}
a & b  \\
- b^* & a^*
\end{pmatrix}\label{eq:ab}
\end{equation}
with some complex numbers $a$ and $b$ such that $|a|^2+|b|^2=1$~\cite{Wigner}.
Hence, any matrix in ${\rm SU}(2)$ can be written as
\begin{equation}
\begin{pmatrix}
w + \imu z & y + \imu x \\
- y + \imu x  &   w - \imu z
\end{pmatrix}
=
w I + \imu ( x X + y Y + z Z )
  \label{eq:Uxyzw}
\end{equation}
with some real numbers $x,y,z$, and $w$ such that
$
w^2+x^2+y^2+z^2=1
$.
Take a real number $\theta$ such that $\cos(\theta/2)=w$
and $\sin(\theta/2)=\sqrt{1-w^2}=\sqrt{x^2+y^2+z^2}$; write $x,y$, and $z$
as $x=-v_x\sin(\theta/2),y=-v_y\sin(\theta/2)$, and
$z=-v_z\sin(\theta/2)$, where $v_x,v_y,v_z\in\SRN$ and $v_x^2+v_y^2+v_z^2=1$.
Thus, using real numbers $\theta,v_x,v_y,v_z\in\SRN$
with $v_x^2+v_y^2+v_z^2=1$,
any matrix in ${\rm SU}(2)$ can be written as
\[
(\cos\mfrac{\theta}{2}) I - \imu (\sin\mfrac{\theta}{2})
(v_x X + v_y Y + v_z Z) ,
\]
which is nothing but $R_{\myv{v}}(\theta)$ in (\ref{eq:Rtheta}).

\subsection{Homomorphism from SU(2) onto SO(3)\label{sssub:homomorphism}}

For $U \in {\rm SU}(2)$, 
we 
denote by $F(U)$ the matrix of the linear transformation on $\SRN^3$ that sends
$(x,y,z)\transp$ to $(x',y',z')\transp$ through
\begin{equation}\label{eq:Uconj}
U(xX+yY+zZ)U^{\dagger}=x'X+y'Y+z'Z .
\end{equation}
Namely, for any $(x,y,z)\transp,(x',y',z')\transp \in \SRN^3$ with (\ref{eq:Uconj}),
\[
\begin{pmatrix}
x' \\ y' \\ z'
\end{pmatrix}
= F(U)
\begin{pmatrix}
x \\ y \\ z
\end{pmatrix} .
\]
We also define 
\begin{equation}
\thDR_{\myv{v}}(\theta):=
F\big(R_{\myv{v}}(\theta)\big), \quad 
\myv{v}\in S^2,\theta\in\SRN.
\end{equation}

\subsection{Generic Orthogonal Axes and Coordinate Axes\label{sssub:lmyz}}
Lemma~\ref{lem:para} can be generalised as follows.
\begin{lemma}\label{lem:paraEulerG}
Let $\myv{l},\myv{m}\in S^2$ be vectors with $\rinnprsp{\myv{l}}{\myv{m}}=0$. Then,
for any $V\in {\rm SU}(2)$,
there exist some 
$\anga,\angc\in\SRN$, and $\angb\in [0,\pi]$ such that
\begin{equation}
V= R_{\myv{m}}(\anga) R_{\myv{l}}(\angb) R_{\myv{m}}(\angc) .
 \label{eq:Eulerlm}
\end{equation}
\end{lemma}

\mynoindent%
{\em Proof.} \ 
Since $F$ is onto ${\rm SO}(3)$,
there exists an element $U \in {\rm SU}(2)$ such that
$\myv{l}=F(U) (0,1,0)\transp \quad\mbox{and}\quad \myv{m}=F(U) (0,0,1)\transp$.%
\footnote{For the sake of constructiveness, 
such an element $U$ is constructed in \refappcomma{app:pr_lm}.}
With this element $U$, 
some $\anga,\angc\in\SRN$, and some $\angb\in [0,\pi]$,
write 
$U^{\dagger} V U = R_{z}(\anga) R_{y}(\angb) R_{z}(\angc)$
in terms of the parameterisation (\ref{eq:uni1}).
Then, since $UR_{z}(\anga)U^{\dagger}=R_{\myv{m}}(\anga)$
$UR_{y}(\angb)U^{\dagger}=R_{\myv{l}}(\angb)$,
and $UR_{z}(\angc)U^{\dagger}=R_{\myv{m}}(\angc)$,
we obtain (\ref{eq:Eulerlm}).
\enproof

We also have Lemma~\ref{lem:transR3}, which is easy but worth recognising.

\begin{lemma}\label{lem:transR3}
Let arbitrary $\kappa, \nu \in \SNN$,
$\myv{u}_1,\dots,\myv{u}_\kappa,\myv{v}_1,\dots,\myv{v}_{\nu} \in S^2$,
and $U\in {\rm SU}(2)$ be given. 
Put 
$\myv{u}_1' = F(U)\myv{u}_1,\dots,\myv{u}_\kappa'=F(U)\myv{u}_\kappa,
\myv{v}_1' = F(U)\myv{v}_1,\dots$, and $\myv{v}_\nu'=F(U)\myv{v}_{\nu}$.
Then, for any
$\theta_1,\dots, \theta_{\kappa},\phi_1,\dots \phi_{\nu} \in \SRN$,
\[
R_{\myv{u}_1}(\theta_1) \cdots R_{\myv{u}_\kappa}(\theta_\kappa)
= R_{\myv{v}_1}(\phi_1) \cdots R_{\myv{v}_{\nu}}(\phi_{\nu})
\]
if and only if (iff)
\[
R_{\myv{u}'_1}(\theta_1) \cdots R_{\myv{u}'_\kappa}(\theta_\kappa)
= R_{\myv{v}'_1}(\phi_1) \cdots R_{\myv{v}'_{\nu}}(\phi_{\nu}) .
\]
\end{lemma}

\mynoindent
{\em Proof.}\/ This readily 
follows from $UR_{\myv{u}_j}(\theta_j)U^{\dagger}=R_{\myv{u}'_j}(\theta_j)$
and $UR_{\myv{v}_j}(\phi_j)U^{\dagger}=R_{\myv{v}'_j}(\phi_j)$.
\enproofwosp

\section{%
The Minimum Numbers of Constituent Rotations and
Optimal Constructions of an Arbitrary Rotation 
\label{ss:const}}

Here
we present the result establishing $N_{\myv{m},\myv{n}}(U)$ with needed definitions. 

\begin{definition}
For $\myv{v}\in S^2$ and
\begin{equation}
U=
\begin{pmatrix}
w + \imu z & y + \imu x \\
- y + \imu x  &   w - \imu z
\end{pmatrix}
=
wI + \imu ( x X + y Y + z Z ) \ \ \in \ {\rm SU}(2)
\label{eq:u123}
\end{equation}
where $w,x,y,z\in \SRN$ are parameters to 
express $U$ uniquely, 
$\myb(\myv{v},U)$ is defined by
\begin{equation}
\myb(\myv{v},U) := \abs{(x,y,z)\myv{v}} .
\label{eq:myb}
\end{equation}
\end{definition}

\begin{definition}\label{def:g}
Functions $f: \SRN^3 \to [0,\pi]$ and $g: \SRN^2 
\times (0,\pi/2] \to \SNN$ are defined by
\begin{align*}
f& (\anga,\angb,\delta):= \\
& 2 \arccos \sqrt{
\cos^2\frac{\angb}{2}\cos^2\frac{\delta}{2}+
\sin^2\frac{\angb}{2}\sin^2\frac{\delta}{2} 
\mbox{}+2\cos\anga
\sin\frac{\angb}{2}\sin\frac{\delta}{2}
\cos\frac{\angb}{2}\cos\frac{\delta}{2} }
\end{align*}
and
\[ 
g(\anga,\angb,\delta) :=
\begin{cases}
\mbox{}\  {\displaystyle 2\Big\lceil \frac{f(\anga,\angb,\delta)}{2\delta} + \frac{1}{2} \Big\rceil }
  & \mbox{if $f(\anga,\angb,\delta) \ge \delta$}\\
\mbox{}\ 4 & \mbox{otherwise.}
\end{cases}
\] 
\end{definition}


\begin{theorem}\label{th:num_rot}
For any $\myv{m},\myv{\varvn}\in S^2$ with 
$\rinnpr{\myv{m}}{\myv{n}} \in [0,1)$,
$\anga,\angc\in\SRN$, and $\angb\in [0,\pi]$, if
\[
\myb(\myv{m},U_{\anga,\angb,\angc}^{\myv{m},\myv{l}}) \ge \myb(\myv{n},U_{\anga,\angb,\angc}^{\myv{m},\myv{l}}) ,
\]
then
\[ 
N_{\myv{m},\myv{n}}\big(F(U_{\anga,\angb,\angc}^{\myv{m},\myv{l}})\big) =
N_{\myv{m},\myv{n}}(U_{\anga,\angb,\angc}^{\myv{m},\myv{l}}) 
= 
\min \Big\{ 2 \Big\lceil \frac{\beta}{2\delta} \Big\rceil +1 , \, 
g(\anga,\angb,\delta) , \,
g(\angc,-\angb,\delta)
 \Big\}
\]
where 
$\delta=\arccos\rinnpr{\myv{m}}{\myv{\varvn}} \in (0,\pi/2]$,
$\myv{l} =\| \myv{m} \times \myv{n} \|^{-1} \myv{m} \times \myv{n}$, and
\[
U_{\anga,\angb,\angc}^{\myv{m},\myv{l}} := R_{\myv{m}}(\anga)R_{\myv{l}}(\angb)
R_{\myv{m}}(\angc) .
\]
\end{theorem}

Note that there is no loss of generality in assuming 
$\myb(\myv{m},U_{\anga,\angb,\angc}^{\myv{m},\myv{l}}) \ge$
$\myb(\myv{n},U_{\anga,\angb,\angc}^{\myv{m},\myv{l}})$,
but also note that $\anga,\angb$ and $\angc$ vary, in general, if $\myv{m}$ and $\myv{n}$ are interchanged.

We give two constructions or decompositions, which will turn out to attain the minimum number 
$N_{\myv{m},\myv{n}}(U_{\anga,\angb,\angc}^{\myv{m},\myv{l}})$ in the theorem.

\begin{proposition}\label{lem:const_odd}
Given arbitrary 
$\myv{m},\myv{\varvn}\in S^2$ with $\rinnpr{\myv{m}}{\myv{n}} \in [0,1)$,
$\anga,\angc\in\SRN$, and $\angb\in [0,\pi]$, 
put
\begin{equation}\label{eq:premise_delta}
\delta=\arccos\rinnpr{\myv{m}}{\myv{\varvn}}\in (0,\pi/2]
\end{equation}
and
\[
\myv{l} =\| \myv{m} \times \myv{n} \|^{-1} \myv{m} \times \myv{n}.
\]
Then, for any $\numK\in\SNN$ and $\angb_1, \dots, \angb_\numK \in (0, 2\delta]$ satisfying
\begin{equation}\label{eq:beta1k}
\angb=\angb_1 + \cdots +\angb_\numK ,
\end{equation}
there exist some $\anga_j,\angc_j,\theta_j\in\SRN$ such that
\begin{equation}\label{eq:pro_RRRR}
R_{\myv{l}}(\angb_j )=
R_{\myv{m}}(-\anga_j) R_{\myv{n}}(\theta_j) 
 R_{\myv{m}}(-\angc_j)
\end{equation}
for $j=1,\dots,\numK$. For these parameters, it holds that
\begin{align}
 R_{\myv{m}}(\anga)&  R_{\myv{l}}(\angb) R_{\myv{m}}(\angc) = \notag\\
& R_{\myv{m}}(\anga-\anga_1)R_{\myv{n}}(\theta_1)
 R_{\myv{m}}(-\angc_1-\anga_2)R_{\myv{n}}(\theta_2) 
 R_{\myv{m}}(-\angc_2-\anga_3)R_{\myv{n}}(\theta_3) \cdots \notag\\
& \cdot R_{\myv{m}}(-\angc_{\numK-1}-\anga_{\numK})R_{\myv{n}}(\theta_{\numK})
 R_{\myv{m}}(-\angc_{\numK}+\angc) .\label{eq:pro_prod}
\end{align}
\end{proposition}

\begin{remark}\label{rem:const}
The least value of $\numK$ such that (\ref{eq:beta1k}) holds for some 
$\angb_1, \dots, \angb_\numK \in (0,2\delta]$ is $\lceil \angb/(2\delta) \rceil $.%
\footnote{
To make the construction explicit, one can set $\angb_j = 2\delta$ for $j \ne \numK$.
The analogous comment applies to the division of $\angb'+\delta$ in Proposition~\ref{lem:const_even}.}
Hence, this proposition gives a decomposition of an arbitrary element 
$U = R_{\myv{m}}(\anga)  R_{\myv{l}}(\angb) R_{\myv{m}}(\angc) \in {\rm SU}(2)$
into the product of 
$2\lceil \angb/(2\delta) \rceil +1$ rotations.%
\footnote{All remarks except Remark~\ref{rem:const}, which needs no proof, 
will be proved in what follows.}
\end{remark}

\begin{remark}\label{rem:const_expl}
For $\mtilde{\angb},\delta\in\SRN$ with $0 \le \mtilde{\angb}/2 \le \delta \le \pi/2$, $\delta \ne 0$,
and $t \in \SRN$, let
\[
H_t(\mtilde{\angb},\delta) :=
\begin{cases}
\mbox{}\ 0 & \mbox{if $\mtilde{\angb}/2 <\delta = \pi/2$}\\
\mbox{}\ t & \mbox{if $\mtilde{\angb}/2 = \delta = \pi/2$}\\
{\displaystyle \arcsin \frac{\tan (\mtilde{\angb}/2) }{\tan \delta}} & \mbox{otherwise.}
\end{cases}
\]
Then, an explicit instance of the set of parameters $\anga_j, \angc_j$, and $\theta_j$
for which 
(\ref{eq:pro_RRRR}) 
holds is given by $(\anga_j,\angc_j,\theta_j)\transp=\myvec{\sigma}_{t_j}(\angb_j,\delta)$,
where
\begin{equation}
\myvec{\sigma}_t(\angb,\delta):=
\begin{pmatrix}
H_t(\angb, \delta) - \slfrac{\pi}{2}\\
H_t(\angb, \delta) + \slfrac{\pi}{2}\\
\mbox{}\ 2 \arcsin \frac{\displaystyle \sin (\angb/2) }{\displaystyle \sin \delta}
\end{pmatrix}
\end{equation}
and $t_j\in\SRN$ can be chosen arbitrarily, $j=1,\dots,\numK$.
[These make (\ref{eq:pro_prod}) hold.]
\end{remark}

\begin{proposition}\label{lem:const_even}
Given any $\myv{m},\myv{\varvn}\in S^2$ with $\rinnpr{\myv{m}}{\myv{n}} \in [0,1)$,
put $\delta=\arccos\rinnpr{\myv{m}}{\myv{\varvn}}\in (0,\pi/2]$ and
$
\myv{l} =\| \myv{m} \times \myv{n} \|^{-1} \myv{m} \times \myv{n}
$.
For an arbitrary $U \in {\rm SU}(2)$,
choose parameters $\anga',\angc'\in \SRN$, and $\angb'\in [0,\pi]$ such that
\begin{equation}\label{eq:const_even}
R_{\myv{l}}(-\delta)U = R_{\myv{m}}(\anga')  R_{\myv{l}}(\angb') R_{\myv{m}}(\angc') .
\end{equation}
Then,
\begin{equation} \label{eq:pro_prod_even0}
U  =  R_{\myv{n}}(\anga')  R_{\myv{l}}(\angb'+\delta) R_{\myv{m}}(\angc')  .
\end{equation}
Furthermore,
for any $\numK' \in \SNN$ and $\angb'_1, \dots, \angb'_{\numK'} \in (0, 2\delta]$ satisfying
\begin{equation}\label{eq:beta1kprime}
\angb'+\delta=\angb'_1 + \cdots +\angb'_{\numK'} ,
\end{equation}
there exist some $\anga'_j, \angc'_j, \theta'_j\in\SRN$ such that
\begin{equation}\label{eq:pro_RRRR_even}
R_{\myv{l}}(\angb'_j )=
R_{\myv{m}}(-\anga'_j) R_{\myv{n}}(\theta'_j) 
 R_{\myv{m}}(-\angc'_j)
\end{equation}
for $j=1,\dots,\numK'$. For these parameters, it holds that
\begin{align}
U = & \
R_{\myv{n}}(\anga') R_{\myv{m}}(-\anga'_1)R_{\myv{n}}(\theta'_1) 
 R_{\myv{m}}(-\angc'_1-\anga'_2)R_{\myv{n}}(\theta'_2) 
 R_{\myv{m}}(-\angc'_2-\anga'_3)R_{\myv{n}}(\theta'_3) \cdots \notag\\
& \ 
 \cdot R_{\myv{m}}(-\angc'_{\numK'-1}-\anga'_{\numK'})R_{\myv{n}}(\theta'_{\numK'})
 R_{\myv{m}}(-\angc'_{\numK'}+\angc') . \label{eq:pro_prod_even}
\end{align}
\end{proposition}

\begin{remark}\label{rem:const_even}
The least value of $\numK'$ such that (\ref{eq:beta1kprime}) holds for some 
$\angb'_1, \dots, \angb'_{\numK'} \in (0,2\delta]$ is 
$\lceil  (\angb' +\delta)/ (2\delta) \rceil
=\lceil  \angb' / (2\delta) + 1/2 \rceil$.
Moreover, if $\angb' \ge \delta$ and $\numK'=\lceil \angb'/(2\delta) + 1/2 \rceil $,
the parameter $\anga'_1$ 
can be chosen 
so that it satisfies $\anga'_1=0$ 
as well as 
(\ref{eq:pro_RRRR_even}) and (\ref{eq:pro_prod_even}).
Hence, when $\angb' \ge \delta$, this proposition and the fact just mentioned give
a decomposition of an arbitrary element 
$U = R_{\myv{n}}(\anga')  R_{\myv{l}}(\angb'+\delta) R_{\myv{m}}(\angc') \in {\rm SU}(2)$
into the product of 
$2 \lceil  \angb' / (2\delta) + 1/2 \rceil$ rotations,
and when $\angb' < \delta$, a decomposition of 
$U$ into the product of 
four rotations.
\end{remark}

\begin{remark}\label{rem:const_expl_even}
An explicit instance of the set of parameters $\anga'_j, \angc'_j$, and $\theta'_j$, $j=1,\dots,\numK'$, for which 
(\ref{eq:pro_RRRR_even}) and (\ref{eq:pro_prod_even}) hold is given by
$(\anga'_j,\angc'_j,\theta'_j)\transp=\myvec{\sigma}_{t_j}(\angb'_j,\delta)$,
where $t_j\in\SRN$ can be chosen arbitrarily, $j=1,\dots,\numK'$.
\end{remark}

\section{%
Limits on Constructions \label{ss:limits}}

In order to bound $N_{\myv{m},\myv{n}}( D )$, etc., from below,
we use the geodesic metric on the unit sphere $S^2$, 
which is denoted by $d$.
Specifically, 
\begin{equation}
d(\myv{u},\myv{v}) := \arccos \rinnpr{\myv{u}}{\myv{v}}
\in [0,\pi]
\end{equation}
for
$\myv{u},\myv{v}\in S^2$. This is the length of
the geodesic connecting $\myv{u}$ and $\myv{v}$ on $S^2$.
We have the following lemma.
[Recall we have put $\thDR_{\myv{v}}(\theta)=
F\big(R_{\myv{v}}(\theta)\big)$.] 

\begin{lemma}\label{lem:ti}
Let $\myv{n},\myv{m}$ be arbitrary 
vectors in $S^2$ with 
$\delta = d(\myv{m},\myv{n}) = \arccos \rinnpr{\myv{m}}{\myv{n}} \in (0,\pi]$.
Then,
for any $\vark \in \SNN$ and $\angg_1,\dots,\angg_{2\vark}\in\SRN$, the following inequalities hold:
\begin{gather}
d(\thDR_{\myv{m}}(\angg_{2\vark-1}) \thDR_{\myv{n}}(\angg_{2\vark-2}) \cdots \thDR_{\myv{m}}(\angg_3)
\thDR_{\myv{n}}(\angg_{2})\thDR_{\myv{m}}(\angg_1)\myv{m}, \myv{m}) \le 2 (\vark-1) \delta,  \label{eq:lem3_3}\\
d(\thDR_{\myv{m}}(\angg_{2\vark-1}) \thDR_{\myv{n}}(\angg_{2\vark-2}) \cdots \thDR_{\myv{m}}(\angg_3)
\thDR_{\myv{n}}(\angg_{2}) \thDR_{\myv{m}}(\angg_1)\myv{m}, \myv{n}) \le (2 \vark -1) \delta, \label{eq:lem3_4}\\
d(\thDR_{\myv{n}}(\angg_{2\vark}) \thDR_{\myv{m}}(\angg_{2\vark-1}) \cdots \thDR_{\myv{m}}(\angg_3)
\thDR_{\myv{n}}(\angg_{2}) \thDR_{\myv{m}}(\angg_1)\myv{m}, \myv{n}) \le (2 \vark -1) \delta , \label{eq:lem3_2} \\
d(\thDR_{\myv{n}}(\angg_{2\vark}) \thDR_{\myv{m}}(\angg_{2\vark-1}) \cdots \thDR_{\myv{m}}(\angg_3)
\thDR_{\myv{n}}(\angg_{2}) \thDR_{\myv{m}}(\angg_1)\myv{m}, \myv{m}) \le 2 \vark \delta. \label{eq:lem3_1} 
\end{gather}
\end{lemma}
This can be shown 
easily by induction on $\vark$
using the triangle inequality for $d$.
In what follows, 
(\ref{eq:lem3_3}) and (\ref{eq:lem3_2}) will be used in the following forms:
\begin{equation} 
2\Big\lceil \frac{d(D\myv{m},\myv{m})}{2\delta} \Big\rceil +1 \le 2 \vark -1 \quad\mbox{and}\quad
2\Big\lceil \frac{d(D'\myv{m},\myv{n})}{2\delta} +\frac{1}{2}\Big\rceil  \le 2 \vark .
\end{equation} 
These bounds hold when $D$ and $D' \in {\rm SO}(3)$ equal the product of $2\vark-1$ rotations and that of $2\vark$ rotations, respectively, in Lemma~\ref{lem:ti} (since $\vark$ is an integer). 
It will turn out that these bounds are tight.

\section{Proof of the Results \label{ss:Proofs}}

\subsection{Structure of the Proof}

Here the structure of the whole proof of the results in this work is described.
Theorem~\ref{th:num_rot} is obtained as a consequence of Lemma~\ref{lem:1} to be presented.
The constructive half of Lemma~\ref{lem:1} is due to Propositions~\ref{lem:const_odd} and \ref{lem:const_even}. 
The other half of Lemma~\ref{lem:1}, related to limits on constructions,
is due to Lemma~\ref{lem:ti}. 
Theorem~\ref{th:Decomp} is derived from Theorem~\ref{th:num_rot} in Appendix~\ref{ss:proofthDecomp}.

\subsection{Proof of Propositions~\ref{lem:const_odd} and \ref{lem:const_even}}

The following lemma is fundamental to the results in this work.

\begin{lemma}\label{lem:EulerG}
For any $\angb,\theta\in\SRN$ and 
for any $\myv{\varu}, \myv{l},\myv{m} \in S^2$
such that $\rinnprsp{\myv{l}}{\myv{m}}=0$,
the following two conditions are equivalent.\vspace{1ex}

\mbox{}\hspace{.5ex}I. There exist some $\anga,\angc\in\SRN$ such that 
\begin{equation}\label{eq:EulerG}
R_{\myv{\varu}}(\theta) 
= R_{\myv{m}}(\anga) R_{\myv{l}}(\angb)
 R_{\myv{m}}(\angc) .
\end{equation}

\mbox{}\hspace{.5ex}II. $\sqrt{1-(\rinnpr{\myv{m}}{\myv{\varu}})^2}\abs{\sin\sfrac{\theta}{2}}
 =\abs{\sin\sfrac{\angb}{2}}$.
\end{lemma}

\noindent
{\em Proof.}  

1)
Take an element $U\in {\rm SU}(2)$ such that
\begin{equation}
\myv{l}=F(U) (0,1,0)\transp \quad\mbox{and}\quad \myv{m}=F(U) (0,0,1)\transp ,
\end{equation}
and put $\myv{v}=(v_x,v_y,v_z)\transp$ for the parameters $v_x,v_y$, and $v_z$ such that
\begin{equation} 
\myv{\varu}= v_x \myv{l} \times \myv{m} + v_y \myv{l} + v_z \myv{m}.\label{eq:u_v}
\end{equation}
Then, owing to Lemma~\ref{lem:transR3}, (\ref{eq:EulerG}) 
holds iff 
\begin{equation}
R_{\myv{v}}(\theta) = R_{z}(\anga) R_{y}(\angb) R_{z}(\angc) .
\label{eq:v_zyz}
\end{equation}

2) 
A direct calculation shows
\begin{eqnarray}
R_{z}(\anga) R_{y}(\angb) R_{z}(\angc)
&=&    \cos\frac{\angb}{2} \cos\frac{\angc+\anga}{2} I
-\imu\sin\frac{\angb}{2} \sin\frac{\angc-\anga}{2} X \notag\\
&&  \mbox{}-\imu\sin\frac{\angb}{2} \cos\frac{\angc-\anga}{2} Y
-\imu\cos\frac{\angb}{2} \sin\frac{\angc+\anga}{2} Z .\label{eq:zyz}
\end{eqnarray}
Hence, (\ref{eq:v_zyz}) is equivalent to
\begin{numcases}
{}
\cos\mfrac{\theta}{2}  =  \cos\mfrac{\angb}{2} 
  \cos\mfrac{\angc+\anga}{2}\label{eq:1}\\
\vargv_x \sin\mfrac{\theta}{2}  =  \sin\mfrac{\angb}{2} 
  \sin\mfrac{\angc-\anga}{2} \label{eq:2}\\
\vargv_y \sin\mfrac{\theta}{2}  =  \sin\mfrac{\angb}{2} 
  \cos\mfrac{\angc-\anga}{2} \label{eq:3}\\
\vargv_z \sin\mfrac{\theta}{2}  =  \cos\mfrac{\angb}{2} 
  \sin\mfrac{\angc + \anga}{2} . \label{eq:4}
\end{numcases}

3) We shall prove I $\Rightarrow$ II.
On each side of (\ref{eq:2}) and (\ref{eq:3}), 
squaring and
summing the resultant pair, we have 
\begin{equation}
\sqrt{1-\vargv_z^2}\abs{\sin\mfrac{\theta}{2}}=\abs{\sin\mfrac{\angb}{2}}.\label{eq:IIorg}
\end{equation}
[Eqs.\ (\ref{eq:1}) and (\ref{eq:4}) also imply (\ref{eq:IIorg}) 
similarly.]
But (\ref{eq:IIorg}) implies II in view of (\ref{eq:u_v}). 

4) Next, we shall prove II $\Rightarrow$ I.

Transforming $(\anga,\angb)$ into $(\eta,\zeta)$, where the two pairs 
are related by
\begin{equation}
\eta=\frac{\angc+\anga}{2} \quad\mbox{and}\quad \zeta=\frac{\angc-\anga}{2},
 \label{eq:5_6}
\end{equation}
we see, from the paragraphs 1) and 2), that
I is equivalent to the following condition:
There exist some $\eta,\zeta\in\SRN$ such that
\begin{numcases}
{}
\cos\mfrac{\theta}{2}  =  \cos\mfrac{\angb}{2} \cos\eta
\label{eq:7} \\
\vargv_x \sin\mfrac{\theta}{2}  =  \sin\mfrac{\angb}{2} \sin\zeta
 \label{eq:8}\\
\vargv_y \sin\mfrac{\theta}{2}   =  \sin\mfrac{\angb}{2} 
  \cos\zeta \label{eq:9}\\
\vargv_z \sin\mfrac{\theta}{2}   =  \cos\mfrac{\angb}{2}
  \sin\eta. \label{eq:10}
\end{numcases}
Hence, 
it is enough to
show that
II implies
the existence of some $\eta,\zeta\in\SRN$ satisfying (\ref{eq:7})--(\ref{eq:10}).

Now suppose $\cos\mfrac{\angb}{2} \ne 0$.
Then, if we show
\begin{equation}
\frac{ \cos^2\mfrac{\theta}{2} }{ \cos^2\mfrac{\angb}{2} } +
\frac{ \vargv_z^2 \sin^2\mfrac{\theta}{2} }{ \cos^2\mfrac{\angb}{2} } = 1 , 
 \label{eq:pr1}
\end{equation}
it will immediately 
imply 
the existence of $\eta$ satisfying (\ref{eq:7}) and (\ref{eq:10}).
From 
II, however, we have (\ref{eq:IIorg}), and hence, 
$(1-\vargv_z^2)\sin^2\mfrac{\theta}{2}=\sin^2\mfrac{\angb}{2}$,
i.e., 
$1-(1-\vargv_z^2)\sin^2\mfrac{\theta}{2}=\cos^2\mfrac{\angb}{2}$,
which is equivalent to (\ref{eq:pr1}) by the assumption 
$\cos\mfrac{\angb}{2} \ne 0$.
If $\cos\mfrac{\angb}{2} = 0$, then $\abs{\sin\mfrac{\angb}{2}}=1$.
This and (\ref{eq:IIorg}) imply
$1-\vargv_z^2=\abs{\sin\mfrac{\theta}{2}}=1$, and hence, 
$\vargv_z=\cos\mfrac{\theta}{2}=0$.
Then, 
(\ref{eq:7}) and (\ref{eq:10}) hold for any choice of $\eta$.

In a similar way, if $\sin\mfrac{\angb}{2} \ne 0$,
\begin{equation}
\frac{ \vargv_x^2 \sin^2\mfrac{\theta}{2} }{ \sin^2\mfrac{\angb}{2} } +
\frac{ \vargv_y^2 \sin^2\mfrac{\theta}{2} }{ \sin^2\mfrac{\angb}{2} } = 1
 \label{eq:pr2}
\end{equation}
will immediately imply  
the existence of $\zeta$ satisfying (\ref{eq:8}) and (\ref{eq:9}).
But (\ref{eq:pr2}) follows again from II or (\ref{eq:IIorg})
since $1-\vargv_z^2=\vargv_x^2+\vargv_y^2$.
If $\sin\mfrac{\angb}{2} = 0$, 
both (\ref{eq:8}) and (\ref{eq:9}) hold for any choice of $\zeta$ similarly.
\enproof

\mynoindent%
{\em Proof of Proposition~\ref{lem:const_odd}.}\/ \hspace*{2.8ex}
Choose a parameter 
$\theta_j$
such that $\abs{\sin(\theta_j/2)}=$ $\sin(\angb_j/2)/\sin\delta$, 
which is possible
by the assumption $\angb_j \in (0, 2\delta]$; then,
it follows from Lemma~\ref{lem:EulerG}
that there exist some $\anga_j,\angc_j\in \SRN$ such that
(\ref{eq:pro_RRRR}), i.e.,
$R_{\myv{l}}(\angb_j )=
R_{\myv{m}}(-\anga_j) R_{\myv{n}}(\theta_j) 
 R_{\myv{m}}(-\angc_j)$ holds,
$j=1,\dots,\numK$.
Inserting these into 
\[
R_{\myv{m}}(\anga)R_{\myv{l}}(\angb)R_{\myv{m}}(\angc)=R_{\myv{m}}(\anga)R_{\myv{l}}(\angb_1) \cdots R_{\myv{l}}(\angb_\numK) R_{\myv{m}}(\angc),
\]
we obtain (\ref{eq:pro_prod}).
\enproof

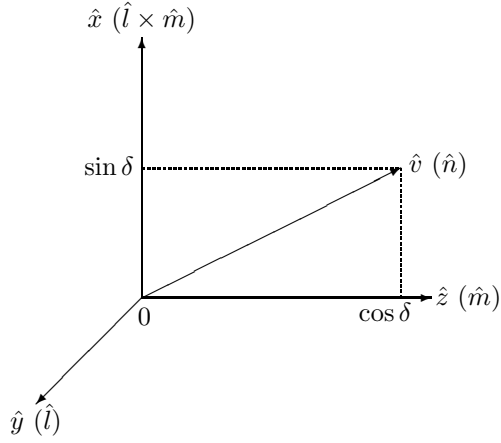
\begin{figure} 
\begin{center}
\begin{picture}(150,165)(-50,-55)
\put(111,0){\makebox(0,0)[l]{$\myv{z}$\ ($\myv{m}$)}} 
\put(0,0){\vector(1,0){109.5}}
\put(0,100){\makebox(0,0)[b]{$\myv{x}$\ ($\myv{l} \times \myv{m}$)}}
\put(0,0){\vector(0,1){98.5}}
\put(1,-4){\makebox(0,0)[t]{$0$}}
\put(0,0){\vector(-1,-1){40}}
\put(-40,-40){\makebox(0,0)[t]{$\myv{y}$\ ($\myv{l}$)}}
\put(0,0){\vector(2,1){98}}
\put(101,50){\makebox(0,0)[l]{$\myv{v}$\ ($\myv{n}$)}}
\multiput(98,0)(0,2){25}{\line(0,1){1}}
\put(92,-2){\makebox(0,0)[t]{$\cos\delta$}}
\multiput(0,49)(2,0){49}{\line(1,0){1}}
\put(-3,50){\makebox(0,0)[r]{$\sin\delta$}}
\end{picture}
\caption{Configuration of $\myv{l},\myv{m}$, and $\myv{n}$ in Propositions~\ref{lem:const_odd} and \ref{lem:const_even}, 
and configuration of $\myv{y}=(0,1,0)\transp$, $\myv{z}=(0,0,1)\transp$, and $\myv{v}$ in arguments around these propositions \label{fig:1}}
\end{center}
\end{figure}

\mynoindent%
{\em Proof of Proposition~\ref{lem:const_even}}.\/
Note $R_{\myv{l}}(\delta)R_{\myv{m}}(\anga')R_{\myv{l}}(-\delta)= R_{\myv{n}}(\anga')$, which is equivalent to
$R_{y}(\delta)R_{z}(\anga')R_{y}(-\delta)=R_{v}(\anga')$, where $\myv{v} = (\sin\delta, 0, \cos\delta)\transp$,
by Lemma~\ref{lem:transR3} (
Figure~\ref{fig:1}) and therefore, can be checked easily by a direct calculation.
Using this equation, 
we can rewrite (\ref{eq:const_even}) as
$U=R_{\myv{n}}(\anga')  R_{\myv{l}}(\angb'+\delta) R_{\myv{m}}(\angc')$, which is (\ref{eq:pro_prod_even0}).
Then, applying to $R_{\myv{l}}(\angb'+\delta) R_{\myv{m}}(\angc')$ the decomposition in Proposition~\ref{lem:const_odd} with $(\anga,\angb,\angc)$ replaced by $(0,\angb'+\delta,\angc')$, 
it readily follows that there exist some $\anga'_j,\angc'_j$, and $\theta'_j \in \SRN$, 
$j=1,\dots,\numK'$, that satisfy the following:
$\abs{\sin(\theta'_j/2)}=\sin(\angb'_j/2)/\sin\delta$ and (\ref{eq:pro_RRRR_even}) for
$j=1,\dots,\numK'$, and
\begin{align}
R_{\myv{l}}(\angb'+\delta)& R_{\myv{m}}(\angc') \notag\\
=& \
 R_{\myv{m}}(-\anga'_1)R_{\myv{n}}(\theta'_1) 
 R_{\myv{m}}(-\angc'_1-\anga'_2)R_{\myv{n}}(\theta'_2) 
 R_{\myv{m}}(-\angc'_2-\anga'_3)R_{\myv{n}}(\theta'_3) \cdots \notag\\
& \
 \cdot R_{\myv{m}}(-\angc'_{\numK'-1}-\anga'_{\numK'})R_{\myv{n}}(\theta'_{\numK'})
 R_{\myv{m}}(-\angc'_{\numK'}+\angc') . 
\end{align}
Thus, we obtain the proposition.
\enproof

Remarks~\ref{rem:const_expl} and \ref{rem:const_expl_even} to these propositions
are proved in \refapp{app:proof_rem}.
The statement on $\anga'_1$ in Remark~\ref{rem:const_even} follows from 
Remark~\ref{rem:const_expl_even} (put $\angb'_1=2\delta$ and $t_1=\pi/2$) or, more directly, from
an equation
$R_{\myv{l}}(2\delta)=R_{\myv{n}}(\pi)R_{\myv{m}}(-\pi)$, which is equivalent to
$R_{y}(2\delta)=R_{v}(\pi)R_{z}(-\pi)$, where $\myv{v}=(\sin\delta,0,\cos\delta)\transp$,
by Lemma~\ref{lem:transR3}.

\subsection{Proof of Theorem~\ref{th:num_rot} \label{sssub:ProofTh2}}

Let $2\SNN -1$ and $2\SNN$ denote the set of odd numbers in $\SNN$ and 
that of even numbers in $\SNN$, respectively.
We define the following for $\myv{m},\myv{\varvn}\in S^2$ with $\abs{\rinnpr{\myv{m}}{\myv{\varvn}}} < 1$:
\begin{align*}
\varM{\myv{m},\myv{n}}{{\rm odd}}(U) := 
\min \{  \varnu \in 2\SNN \! - \! 1 \mid \, &
\exists V_1, V_3, \dots,V_{\varnu}\in \cR_{\myv{m}},\\
& \exists V_2, V_4, \dots,V_{\varnu-1}\in \cR_{\myv{n}}, \
 U=  V_{\varnu}V_{\varnu-1} \cdots V_1\} ,\\
\varM{\myv{m},\myv{n}}{\rm even}(U) :=
\min \{   \varnu \in 2\SNN \mid \, &
\exists V_1, V_3, \dots,V_{\varnu-1}\in \cR_{\myv{m}},\\
& \exists V_2, V_4, \dots,V_{\varnu}\in \cR_{\myv{n}}, \
 U=  V_{\varnu}V_{\varnu-1} \cdots V_1\} ,\\
\varMcomb{\myv{m},\myv{n}}(U) :=
\min \{  \varM{\myv{m},\myv{n}}{{\rm odd}}(U), & \ \varM{\myv{m},\myv{n}}{\rm even}(U) \} 
\end{align*}
for $U\in {\rm SU}(2)$;
\begin{align*}
\mbox{}\
\varM{\myv{m},\myv{n}}{\rm odd}(D) := 
\min \{  \varnu \in 2\SNN \! - \! 1 \mid \, &
\exists \genD_1, \genD_3, \dots,\genD_{\varnu}\in \hat{\cR}_{\myv{m}},\\
& \mbox{}\!\! \exists \genD_2, \genD_4, \dots,\genD_{\varnu-1}\in \hat{\cR}_{\myv{n}}, \
D=  \genD_{\varnu}\genD_{\varnu-1} \cdots \genD_1\} ,\\
\varM{\myv{m},\myv{n}}{\rm even}(D) :=
\min \{  \varnu \in 2\SNN \mid \, &
\exists \genD_1, \genD_3, \dots,\genD_{\varnu-1}\in \hat{\cR}_{\myv{m}},\\
& \exists \genD_2, \genD_4, \dots,\genD_{\varnu}\in \hat{\cR}_{\myv{n}}, \
 D=  \genD_{\varnu}\genD_{\varnu-1} \cdots \genD_1\} ,\\
\varMcomb{\myv{m},\myv{n}}(D) :=
\min \{ \varM{\myv{m},\myv{n}}{\rm odd}(D), & \ \varM{\myv{m},\myv{n}}{\rm even}(D) \} 
\end{align*}
for $D\in {\rm SO}(3)$.
The following lemma largely solves the issue of determining 
the optimal number $N_{\myv{m},\myv{n}}(U)$.

\begin{lemma}\label{lem:1}
Let $\myv{m},\myv{n}$, $\myv{l}$, and $\delta$
be as in Theorem~\ref{th:num_rot}.
Then, for any $\anga,\angc\in\SRN$, and $\angb\in [0,\pi]$,
\begin{equation}
\varM{\myv{m},\myv{n}}{\rm odd}\big(F(U_{\anga,\angb,\angc}^{\myv{m},\myv{l}})\big) = 
\varM{\myv{m},\myv{n}}{\rm odd}(U_{\anga,\angb,\angc}^{\myv{m},\myv{l}}) = 
2 \Big\lceil \frac{\beta}{2\delta} \Big\rceil +1 \label{eq:lem1_1}
\end{equation}
and
\begin{equation}
\varM{\myv{m},\myv{n}}{\rm even}\big(F(U_{\anga,\angb,\angc}^{\myv{m},\myv{l}})\big) = 
\varM{\myv{m},\myv{n}}{\rm even}(U_{\anga,\angb,\angc}^{\myv{m},\myv{l}}) = 
g(\anga,\angb,\delta) \label{eq:lem1_2}
\end{equation}
where
$U_{\anga,\angb,\angc}^{\myv{m},\myv{l}}$ 
is as defined in Theorem~\ref{th:num_rot}.
\end{lemma}
\begin{corollary}\label{coro:1}
Let $\myv{m},\myv{n}$, $\myv{l}$, and $\delta$
be as in Theorem~\ref{th:num_rot}.
Then, for any $\anga,\angc\in\SRN$, and $\angb\in [0,\pi]$,
\begin{gather}\label{eq:2:DUP_SEP10}
\varMcomb{\myv{m},\myv{n}}\big(F(U_{\anga,\angb,\angc}^{\myv{m},\myv{l}})\big) 
=
\varMcomb{\myv{m},\myv{n}}(U_{\anga,\angb,\angc}^{\myv{m},\myv{l}}) 
= \min \Big\{ 2 \Big\lceil \frac{\beta}{2\delta} \Big\rceil +1 , \,
g(\anga,\angb,\delta) \Big\}.
\end{gather}
\end{corollary}

\mynoindent%
{\em Proof}.\/ 
In the case where $\angb=0$,
since $\varM{\myv{m},\myv{n}}{\rm odd}(U_{\anga,\angb,\angc}^{\myv{m},\myv{l}}) = 1$
and 
$\varM{\myv{m},\myv{n}}{\rm even}(U_{\anga,\angb,\angc}^{\myv{m},\myv{l}}) = 2$,
(\ref{eq:lem1_1}) and (\ref{eq:lem1_2}) are trivially true.
We shall prove the statement for $\angb>0$.

To establish (\ref{eq:lem1_1}), we shall show the first and third inequalities in
\begin{equation}
2 \Big\lceil \frac{\beta}{2\delta} \Big\rceil +1 
\le 
\varM{\myv{m},\myv{n}}{\rm odd}\big(F(U_{\anga,\angb,\angc}^{\myv{m},\myv{l}})\big) 
\le
\varM{\myv{m},\myv{n}}{\rm odd}(U_{\anga,\angb,\angc}^{\myv{m},\myv{l}}) 
\le
2 \Big\lceil \frac{\beta}{2\delta} \Big\rceil +1 
\label{eq:lem1_1_comb}
\end{equation}
while the second inequality 
trivially follows from the definition of $\varM{\myv{m},\myv{n}}{\rm odd}$.

Note first that Remark~\ref{rem:const} to Proposition~\ref{lem:const_odd} immediately implies
the third inequality in (\ref{eq:lem1_1_comb}).
To prove the first 
inequality,
assume 
\begin{equation}
F(U_{\anga,\angb,\angc}^{\myv{m},\myv{l}})= \genD_{\varnu}\genD_{\varnu-1}\cdots \genD_1
\label{eq:pr_lem1_1}
\end{equation}
for some $j=2k-1$ with $k\in\SNN$,
where $\genD_{\nu}\in\hat{\cR}_{\myv{m}}$ if $\nu$ is odd
and $\genD_{\nu}\in\hat{\cR}_{\myv{n}}$ otherwise.

We shall evaluate $d(F(U_{\anga,\angb,\angc}^{\myv{m},\myv{l}} ) \myv{m},\myv{m})= 
d(\genD_{\varnu}\genD_{\varnu-1}\cdots \genD_1 \myv{m},\myv{m})$. \hspace*{1ex} Noting that 
$d(F(U_{\anga,\angb,\angc}^{\myv{m},\myv{l}})\myv{m},\myv{m})=\angb$, we have
$\angb \le 2(k-1)\delta$ by (\ref{eq:lem3_3}) of Lemma~\ref{lem:ti}. 
This implies $\lceil \angb/(2\delta) \rceil \le k-1$, and therefore, 
\begin{equation}\label{eq:oddlb}
2\Big\lceil \frac{\angb}{2\delta} \Big\rceil +1 \le 2k-1=j.
\end{equation}
From this bound,
we have the first inequality in
(\ref{eq:lem1_1_comb}),
and hence, (\ref{eq:lem1_1}).

To establish (\ref{eq:lem1_2}), we shall first treat the major case where $f(\anga,\angb,\delta) \ge \delta$.
Recalling that $g(\anga,\angb,\delta) = 2 \lceil f(\anga,\angb,\delta)/(2\delta) +1/2 \rceil$ in this case, we shall show the first and third inequalities in
\begin{equation}\label{eq:lem1_2_comb}
2 \Big\lceil \frac{f(\anga,\angb,\delta)}{2\delta} + \frac{1}{2} \Big\rceil 
\le
\varM{\myv{m},\myv{n}}{\rm even}\big(F(U_{\anga,\angb,\angc}^{\myv{m},\myv{l}})\big)
\le
\varM{\myv{m},\myv{n}}{\rm even}(U_{\anga,\angb,\angc}^{\myv{m},\myv{l}}) \le
2 \Big\lceil \frac{f(\anga,\angb,\delta)}{2\delta} + \frac{1}{2} \Big\rceil 
\end{equation}
while the second inequality holds trivially.

Note that Remark~\ref{rem:const_even} to Proposition~\ref{lem:const_even} will imply
the third inequality
upon showing that $\angb'$ in Proposition~\ref{lem:const_even} satisfies $\angb'=f(\anga,\angb,\delta)$ 
when $U=U_{\anga,\angb,\angc}^{\myv{m},\myv{l}}$.
To see $\angb'=f(\anga,\angb,\delta)$, rewrite (\ref{eq:const_even}), using Lemma~\ref{lem:transR3}, as
\begin{equation}
R_{y}(-\delta)R_{z}(\anga)R_{y}(\angb)R_{z}(\angc) = R_{z}(\anga')  R_{y}(\angb') R_{z}(\angc') .
\end{equation}
Then, a direct calculation shows the absolute value of the $(1,1)$-entry of the left-hand side equals 
\[
\sqrt{
\cos^2\frac{\angb}{2}\cos^2\frac{\delta}{2}+
\sin^2\frac{\angb}{2}\sin^2\frac{\delta}{2} 
\mbox{}+2\cos\anga
\sin\frac{\angb}{2}\sin\frac{\delta}{2}
\cos\frac{\angb}{2}\cos\frac{\delta}{2} }.
\]
This shows $\angb'=f(\anga,\angb,\delta)$ in view of (\ref{eq:uni1}).

To prove the first inequality in (\ref{eq:lem1_2_comb}) 
assume 
(\ref{eq:pr_lem1_1}) holds 
for some $j=2k$ with $k\in\SNN$,
where $\genD_{\nu}\in\hat{\cR}_{\myv{m}}$ if $\nu$ is odd
and $\genD_{\nu}\in\hat{\cR}_{\myv{n}}$ otherwise.
Note that $\myv{n}=R_{\myv{l}}(\delta)\myv{m}$ and hence, for $U=R_{\myv{n}}(\anga')R_{\myv{l}}(\angb'+\delta)R_{\myv{m}}(\angc')$ in Proposition~\ref{lem:const_even},
\[ 
d(U\myv{m},\myv{n})=d(R_{\myv{l}}(\angb'+\delta)\myv{m},\myv{n})=d(R_{\myv{l}}(\angb'+\delta)\myv{m},R_{\myv{l}}(\delta)\myv{m})=(\angb'+\delta)-\delta=\angb'.
\]
Then, we have
$\angb' \le (2 k -1) \delta$ by (\ref{eq:lem3_2}) of Lemma~\ref{lem:ti}. 
This implies $\lceil (\angb'+\delta)/(2\delta) \rceil \le k$, and therefore, 
\begin{equation}\label{eq:evenlb}
2\Big\lceil \frac{\angb'+\delta}{2\delta} \Big\rceil \le 2k=j.
\end{equation}
From this bound,
we have 
the first inequality in (\ref{eq:lem1_2_comb}) 
and hence, the equality among all sides of (\ref{eq:lem1_2_comb}).
This shows (\ref{eq:lem1_2})  in the case where $f(\anga,\angb,\delta) \ge \delta$. The proof of (\ref{eq:lem1_2}) in the other case is given in \refapp{app:case2}.
This completes the proof of the lemma. 
The proved lemma immediately implies the corollary.
\enproof

\mynoindent%
{\em Proof of Theorem~\ref{th:num_rot}.}\/
Note that for any $U\in {\rm SU}(2)$,
\[ 
N_{\myv{m},\myv{n}}(U)
= \min \{ 
\varM{\myv{m},\myv{n}}{\rm odd}(U),
\varM{\myv{m},\myv{n}}{\rm even}(U),
\varM{\myv{n},\myv{m}}{\rm odd}(U),
\varM{\myv{n},\myv{m}}{\rm even}(U)
\} ,
\]
and we can write $U$ in terms of three parametric expressions:
\begin{equation*}
U=R_{\myv{u}}(\theta)=U_{\anga,\angb,\angc}^{\myv{m},\myv{l}}=
U_{\tilde{\anga},\tilde{\angb},\tilde{\angc}}^{\myv{n},-\myv{l}} 
\end{equation*}
where $\angb,\tilde{\angb} \in [0, \pi]$, $\anga,\angc,\tilde{\anga},\tilde{\angc},\theta \in \SRN$,
and $\myv{u}\in S^2$.
Then, we have
\[
\frac{\beta}{2}=\arcsin \bigg[ \sqrt{1-(\rinnpr{\myv{m}}{\myv{u}})^2}\Big|\sin\frac{\theta}{2}\Big| \bigg]
\quad\mbox{and}\quad 
\frac{\tilde{\beta}}{2}=\arcsin \bigg[ \sqrt{1-(\rinnpr{\myv{n}}{\myv{u}})^2}\Big|\sin\frac{\theta}{2}\Big| \bigg]
\]
owing to Lemma~\ref{lem:EulerG}, and hence, 
\[
\varM{\myv{m},\myv{n}}{\rm odd}(U)=
2\Big\lceil \frac{\arcsin \sqrt{1-(\rinnpr{\myv{m}}{\myv{u}})^2}\abs{\sin\mfrac{\theta}{2}}}{\delta} \Big\rceil +1
\]
and
\[
\varM{\myv{n},\myv{m}}{\rm odd}(U)=
2\Big\lceil \frac{\arcsin \sqrt{1-(\rinnpr{\myv{n}}{\myv{u}})^2}\abs{\sin\mfrac{\theta}{2}}}{\delta} \Big\rceil +1 
\]
owing to Lemma~\ref{lem:1}. Then, if $\abs{\rinnpr{\myv{m}}{\myv{u}}} \ge \abs{\rinnpr{\myv{n}}{\myv{u}}}$
whenever $\sin (\theta/2) \ne 0$, which implies 
$\varM{\myv{m},\myv{n}}{\rm odd}(U) \le \varM{\myv{n},\myv{m}}{\rm odd}(U)$, we shall have
\begin{eqnarray}
N_{\myv{m},\myv{n}}(U)
&= &\min \{ 
\varM{\myv{m},\myv{n}}{\rm odd}(U),
\varM{\myv{m},\myv{n}}{\rm even}(U),
\varM{\myv{n},\myv{m}}{\rm even}(U)
\} \notag\\
&= &\min \Big\{ 
2 \Big\lceil \frac{\beta}{2\delta} \Big\rceil +1,
g(\anga,\angb,\delta),
\varM{\myv{n},\myv{m}}{\rm even}(U)
 \Big\} \label{eq:lem1pre}
\end{eqnarray}
for $U=U_{\anga,\angb,\angc}^{\myv{m},\myv{l}}$. 
But $[\sin (\theta/2) \ne 0 \rightarrow \abs{\rinnpr{\myv{m}}{\myv{u}}} \ge \abs{\rinnpr{\myv{n}}{\myv{u}}}]$ follows from 
$\myb(\myv{m},U_{\anga,\angb,\angc}^{\myv{m},\myv{l}}) \ge \myb(\myv{n},U_{\anga,\angb,\angc}^{\myv{m},\myv{l}})$ 
by the definition of $\myb$.
[This is because writing $U$ in (\ref{eq:u123}) as $U=R_{\myv{u}}(\theta)$,
$\theta\in\SRN$, $\myv{u}\in S^2$, results in
$-\sin(\theta/2) \myv{u} = (x,y,z)\transp$ as in Section~\refroyalsub{ss:prel}{sssub:para},
whereby $\myb(\myv{v},U)=\abs{\sin(\theta/2)} \abs{ \rinnpr{\myv{u}}{\myv{v}} }$.]
Hence, 
we have (\ref{eq:lem1pre}).

A short additional argument (\refappcomma{app:tech}) shows
\begin{equation}
\varM{\myv{n},\myv{m}}{\rm even}(U_{\anga,\angb,\angc}^{\myv{m},\myv{l}})
= g(\angc,-\angb,\delta), \label{eq:a_even}
\end{equation}
and therefore, 
\[
N_{\myv{m},\myv{n}}(U_{\anga,\angb,\angc}^{\myv{m},\myv{l}}) 
= 
\min \Big\{ 2 \Big\lceil \frac{\beta}{2\delta} \Big\rceil +1 , \, 
g(\anga,\angb,\delta) , \,
g(\angc,-\angb,\delta)
 \Big\}.
\]
Finally, from Corollary~\ref{coro:1} or from the argument in \refappcomma{app:N_mnUandD}, 
it readily follows that
$ 
N_{\myv{m},\myv{n}}\big(F(U_{\anga,\angb,\angc}^{\myv{m},\myv{l}})\big) =
N_{\myv{m},\myv{n}}(U_{\anga,\angb,\angc}^{\myv{m},\myv{l}})
$. 
Hence, we obtain
the theorem.
\enproof

From the viewpoint of construction, we summarise the (most directly) suggested way to obtain an optimal construction of a given element $U\in {\rm SU}(2)$,
where we assume $\delta=\arccos\rinnpr{\myv{m}}{\myv{n}}\in (0,\pi/2]$ 
without loss of generality. 
If $\myb(\myv{m},U)$ $\ge \myb(\myv{n},U)$, 
choose a construction
that attains the minimum in (\ref{eq:lem1pre}). The construction is 
among that of Proposition~\ref{lem:const_odd}, that of Proposition~\ref{lem:const_even}, and
that of Proposition~\ref{lem:const_even} applied to $U^{\dagger}$ in place of $U$
[note $U^{\dagger}=R_{\myv{u}_1}(\phi_1) \cdots R_{\myv{u}_j}(\phi_j)$ implies $U=R_{\myv{u}_j}(-\phi_j)\cdots R_{\myv{u}_1}(-\phi_1)$].
If $\myb(\myv{m},U) < \myb(\myv{n},U)$,
interchanging $\myv{m}$ and $\myv{n}$, apply the construction just described.%
\footnote{One (seemingly difficult) issue arises: 
Determine all optimal decompositions of an arbitrarily fixed rotation.
Note that in Propositions~\ref{lem:const_odd} and \ref{lem:const_even}
and their proofs,
any solution for 
$R_{\myv{n}}(\theta) = R_{\myv{m}}(\anga)R_{\myv{l}}(\angb ) R_{\myv{m}}(\angc)$ 
can be used (see Corollary~\ref{coro:8} in \refapp{app:proof_rem} for explicit solutions,
among which one is chosen to be used in Remarks~\ref{rem:const_expl} and \ref{rem:const_expl_even}).}

\section{Conclusion \label{ss:conc}}

This work has established
the least value $N_{\myv{m},\myv{n}}(U)$
of a positive integer $k$ such that $U$ can be decomposed into the product of $k$ rotations about either $\myv{m}$ or $\myv{\varvn}$
for an arbitrarily fixed 
element $U$ in ${\rm SU(2)}$, or in ${\rm SO}(3)$, where $\myv{m},\myv{n} \in S^2$
are arbitrary 
real unit vectors with $\abs{\rinnpr{\myv{m}}{\myv{n}}}<1$.

\section*{Acknowledgments}
This work was supported by SCOPE (Ministry of Internal Affairs and Communications), and by Japan Society for the Promotion of Science KAKENHI Grant numbers 22540150
and 21244007.

\appendix
\section*{Appendices} 

\sectionapp{%
Element in SU(2) Associated with $\myv{l}$ and $\myv{m}$
\label{app:pr_lm}}

Our goal here is to prove (in a constructive manner) that
for any pair of vectors
$\myv{l},\myv{m} \in S^2$ with 
$\rinnprsp{\myv{l}}{\myv{m}}=0$,
there exists some element $U\in{\rm SU}(2)$ such that 
$\myv{l}=F(U)(0,1,0)\transp$ and $\myv{m}=F(U)(0,0,1)\transp$. 
Expressing $U$ as 
$
U=R_z(\tilde{\anga})R_y(\tilde{\angb})R_z(\tilde{\angc})
$,
we shall specify desired $\tilde{\anga},\tilde{\angb}$, and $\tilde{\angc}$.
By a direct calculation with
\begin{equation*}
\thDR_y(\theta) =
\begin{pmatrix}
\cos\theta & 0 & \sin\theta  \\
0 & 1 & 0 \\
-\sin\theta & 0 & \cos\theta 
\end{pmatrix} 
\quad\mbox{and}\quad
\thDR_z(\theta)
=
\begin{pmatrix}
\cos\theta & -\sin\theta & 0 \\
\sin\theta & \cos\theta & 0 \\
0 & 0 & 1
\end{pmatrix}  
\end{equation*}
where $\thDR_y(\theta):=F\big(R_y(\theta)\big)$ and $\thDR_z(\theta):=F\big(R_z(\theta)\big)$,
we have $F(U)(0,0,1)\transp=(\cos\tilde{\anga} \sin\tilde{\angb}, \sin\tilde{\anga} \sin\tilde{\angb}, \cos\tilde{\angb} )\transp$. 
On the other hand,
the condition 
$\myv{l}=F(U)(0,1,0)\transp$ is equivalent to
$
\thDR_y(-\tilde{\angb})\thDR_z(-\tilde{\anga})\myv{l}=\thDR_z(\tilde{\angc})(0,1,0)\transp
$,
i.e.,
\begin{equation}
\begin{pmatrix}
\cos\tilde{\angb} \cos\tilde{\anga} & \cos\tilde{\angb} \sin\tilde{\anga} & -\sin\tilde{\angb}\\
-\sin\tilde{\anga} & \cos\tilde{\anga} & 0\\
\cos\tilde{\anga} \sin\tilde{\angb} & \sin\tilde{\anga} \sin\tilde{\angb} & \cos\tilde{\angb}
\end{pmatrix}
\myv{l} =
\begin{pmatrix}
-\sin\tilde{\angc}\\
\cos\tilde{\angc} \\
0
\end{pmatrix} .\label{eq:app2}
\end{equation}

Hence, 
choosing parameters $\tilde{\anga}$ and $\tilde{\angb}$ such that $(\cos\tilde{\anga} \sin\tilde{\angb}, \sin\tilde{\anga} \sin\tilde{\angb}$, 
$\cos\tilde{\angb} )\transp=\myv{m}$,
cf.\ spherical coordinates,
and $\tilde{\angc}$ that satisfies (\ref{eq:app2}),
we have a desired element $U=R_z(\tilde{\anga})R_y(\tilde{\angb})R_z(\tilde{\angc})$ such that
$\myv{l}=F(U)(0,1,0)\transp$ and $\myv{m}=F(U)(0,0,1)\transp$. 

\sectionapp{Details on Angles in Propositions~\ref{lem:const_odd} and \ref{lem:const_even} \label{app:proof_rem}}

Examining the proof of Lemma~\ref{lem:EulerG},
we can be specific about $\anga$ and $\angc$
to have the following lemma and corollary. \hspace{1ex} 
In particular, the corollary gives a sufficient condition, 
(i), and two necessary conditions,
(ii) and (iii), for  
$R_{\myv{n}}(\theta) = R_{\myv{m}}(\anga) R_{\myv{l}}(\angb) R_{\myv{m}}(\angc)$, 
where $\myv{l},\myv{m}$, and $\myv{n}$ are set as in
Propositions~\ref{lem:const_odd} and \ref{lem:const_even}.
Remarks~\ref{rem:const_expl} and \ref{rem:const_expl_even}
will be clear from (i). 
Later, (ii) and (iii)
will be used in \refapps{app:case2}{app:tech}, respectively, 
though the use of them is not mandatory.

\begin{lemma}\label{lem:7}
For any $\theta,\anga,\angb,\angc\in\SRN$, and $\myv{n}, \myv{l},\myv{m} \in S^2$ such that $\rinnprsp{\myv{l}}{\myv{m}}=0$,
\begin{equation} \label{eq:EulerGn}
R_{\myv{n}}(\theta) 
= R_{\myv{m}}(\anga) R_{\myv{l}}(\angb)
 R_{\myv{m}}(\angc) 
\end{equation}
holds iff the following conditions hold: 
\begin{equation}
\cos\morfrac{\angc+\anga}{2} = \frac{ \cos\mfrac{\theta}{2} }{ 
  \cos\mfrac{\angb}{2} } \quad\mbox{and}\quad
\sin\morfrac{\angc+\anga}{2} = \frac{ \rinnpr{\myv{m}}{\myv{\varvn}} \sin\mfrac{\theta}{2} }{ 
  \cos\mfrac{\angb}{2} } \label{eq:lemetazeta1}
\end{equation}
whenever $\cos \mfrac{\angb}{2} \ne 0$,
\begin{equation}
 \sin\morfrac{\angc-\anga}{2} = \frac{\rinnpr{(\myv{l}\times\myv{m})}{\myv{\varvn}} \sin\mfrac{\theta}{2}}{
  \sin\mfrac{\angb}{2} } \quad\mbox{and}\quad
\cos\morfrac{\angc-\anga}{2} = \frac{\rinnprsp{\myv{l}}{\myv{\varvn}} \sin\mfrac{\theta}{2}}{
 \sin\mfrac{\angb}{2} }  \label{eq:lemetazeta2}
\end{equation}
whenever $\sin \mfrac{\angb}{2} \ne 0$,
and
\begin{equation} \label{eq:lemetheta}
\sqrt{1-(\rinnpr{\myv{m}}{\myv{n}}})^2 \abs{\sin\mfrac{\theta}{2}} = \abs{\sin\mfrac{\angb}{2}} .
\end{equation}
\end{lemma}

\begin{corollary}\label{coro:8}
Given any $\delta\in (0,\pi/2]$ and $\myv{l}, \myv{m}\in S^2$
such that
$\rinnprsp{\myv{l}}{\myv{m}}=0$, 
put 
\begin{equation}
\myv{n}=(\sin\delta) \myv{l}\times\myv{m} + (\cos\delta) \myv{m} .
\end{equation}
Then, (i) for any $\theta,\anga,\angc\in\SRN$, and $\angb\in [0,\pi]$,
(\ref{eq:EulerGn})
holds
if 
\begin{equation*}
\angb \le 2 \delta
\end{equation*}
and
there exists some $t\in \SRN$ such that (recall $H_t$ is defined in Remark~\ref{rem:const_expl})
\begin{equation}\label{eq:coro8para}
\begin{pmatrix}
\anga \\
\angc \\
\theta 
\end{pmatrix}
=\pm
\begin{pmatrix}
H_t(\angb, \delta) - \slfrac{\pi}{2}\\
H_t(\angb, \delta) + \slfrac{\pi}{2}\\
\mbox{}\ 2 \arcsin \frac{\displaystyle \sin (\angb/2) }{\displaystyle \sin \delta} 
\end{pmatrix}
\mbox{}\ \mbox{or}\
\begin{pmatrix}
\anga \\
\angc \\
\theta 
\end{pmatrix}
=\pm
\begin{pmatrix}
-H_t(\angb, \delta) + \slfrac{\pi}{2}\\
-H_t(\angb, \delta) + \slfrac{3\pi}{2}\\
\mbox{}\ 2\pi -2 \arcsin \frac{\displaystyle \sin (\angb/2) }{\displaystyle \sin \delta}
\end{pmatrix} ;
\end{equation}
(ii) for any $\anga\in\SRN$ and $\angb\in (0,\pi]$,
if (\ref{eq:EulerGn}) holds for some $\theta,\angc \in \SRN$,
then $\angb \le 2\delta$ and
there exist some $j\in\SINT$ and $t\in \SRN$ such that\/%
\footnote{Here $w=\pm x \pm y +z$ means $w \in \{ x+y+z,x-y+z,-x+y+z,-x-y+z \}$.}
\begin{equation*} 
\anga =  \pm H_t(\angb, \delta) \pm \slfrac{\pi}{2} + \pi j ;
\end{equation*}
(iii) for any $\angc\in\SRN$ and $\angb\in (0,\pi]$,
if (\ref{eq:EulerGn}) holds for some $\theta,\anga \in \SRN$, 
then $\angb \le 2\delta$ and
there exist some $j\in\SINT$ and $t\in \SRN$ such that  
\begin{equation*} 
\angc =  \pm H_t(\angb, \delta) \pm \slfrac{\pi}{2} + \pi j .
\end{equation*}
\end{corollary}

\mynoindent%
{\em Proof}.\/
Set $\myv{v}=(v_x,v_y,v_z)\transp$ with
\[
v_x = 
\rinnpr{(\myv{l}\times\myv{m})}{\myv{n}}, \quad
v_y=
\rinnprsp{\myv{l}}{\myv{n}}, \quad \mbox{and}\quad
v_z=
\rinnpr{\myv{m}}{\myv{n}}.
\]
Then,
according to the paragraphs 1) and 2) in 
the proof of Lemma~\ref{lem:EulerG}, for any $\theta,\anga,\angb,\angc\in\SRN$,
(\ref{eq:EulerGn}) holds
iff (\ref{eq:1})--(\ref{eq:4}) hold. 
But 
(\ref{eq:1})--(\ref{eq:4}) hold iff
(\ref{eq:lemetheta}), [$\cos \mfrac{\angb}{2} \ne 0 \rightarrow (\mbox{\ref{eq:lemetazeta1}})$], 
and [$\sin \mfrac{\angb}{2} \ne 0 \rightarrow (\mbox{\ref{eq:lemetazeta2}})$] hold.
This completes the proof of the lemma.

To see the corollary, (recall 
Figure~\ref{fig:1} and) note 
\[
\rinnpr{(\myv{l}\times\myv{m})}{\myv{n}}
= \sin\delta , \quad
\rinnprsp{\myv{l}}{\myv{n}}
= 0,
\quad \mbox{and}\quad
\rinnpr{\myv{m}}{\myv{n}}
= \cos\delta .
\]
Then, 
(\ref{eq:lemetheta}), [$\cos \mfrac{\angb}{2} \ne 0 \rightarrow (\mbox{\ref{eq:lemetazeta1}})$], 
and [$\sin \mfrac{\angb}{2} \ne 0 \rightarrow (\mbox{\ref{eq:lemetazeta2}})$] 
hold
if the following two conditions are satisfied: (a) $\angb \le 2\delta$; (b)
\begin{equation*}
\begin{cases}
\mbox{}\ \frac{\angc+\anga}{2} = \arcsin \frac{\mydisplaystyle \tan (\angb/2) }{\mydisplaystyle \tan \delta} \\
\mbox{}\ \frac{\angc-\anga}{2} = \frac{\pi}{2}\\
\mbox{}\ \theta = 2 \arcsin \frac{\mydisplaystyle \sin (\angb/2) }{\mydisplaystyle \sin \delta} 
\end{cases}
\quad \mbox{or} \quad
\begin{cases}
\mbox{}\ 
\frac{\angc+\anga}{2} = 
 \pi -\arcsin \frac{\mydisplaystyle \tan (\angb/2) }{\mydisplaystyle \tan \delta}\\
\mbox{}\ \frac{\angc-\anga}{2} = \frac{\pi}{2}\\
\mbox{}\ \theta = 2\pi - 2 \arcsin \frac{\mydisplaystyle \sin (\angb/2) }{\mydisplaystyle \sin \delta} 
\end{cases}
\end{equation*}
unless $\angb/2=\delta = \pi/2$,%
\footnote{$\frac{\tan (\angb/2) }{\tan \delta}$ should be understood as $0$ if $\angb/2 <\delta = \pi/2$.\label{fn:8}}
and \begin{equation*}
\begin{cases}
\mbox{}\ \frac{\angc+\anga}{2} = s \\
\mbox{}\ \frac{\angc-\anga}{2} = \frac{\pi}{2} \\
\mbox{}\ \theta=\angb
\end{cases} 
\end{equation*}
for some $s\in \SRN$ if $\angb/2=\delta = \pi/2$. This readily gives two solutions for (\ref{eq:EulerGn}).
Rewriting these solutions with $H_t$ and
checking that flipping the signs of the solutions gives other solutions,
we obtain (i).
Showing (ii) and (iii) is as easy as showing (i).
\hspace*{1ex}\mbox{} 
\enproofwosp

\sectionapp{Proofs of (\ref{eq:lem1_2}) in the Case 
$f(\anga,\angb,\delta) < \delta$
 \label{app:case2}}

\mynoindent%
{\em Proof 1.}\/
Proposition~\ref{lem:const_even} and
Remark~\ref{rem:const_even} show $\varM{\myv{m},\myv{n}}{\rm even}(U_{\anga,\angb,\angc}^{\myv{m},\myv{l}}) \le 4$, i.e.,
either 
$\varM{\myv{m},\myv{n}}{\rm even}(U_{\anga,\angb,\angc}^{\myv{m},\myv{l}}) = 2$ or $\varM{\myv{m},\myv{n}}{\rm even}(U_{\anga,\angb,\angc}^{\myv{m},\myv{l}}) = 4$.
We also have $\varM{\myv{m},\myv{n}}{\rm even}\big(F(U)\big)=\varM{\myv{m},\myv{n}}{\rm even}(U)$ for any $U \in {\rm SU}(2)$ \refappparen{app:N_mnUandD}.
Hence, all we need to show is that
\begin{equation}\label{eq:case2_1}
\exists \theta,\phi\in\SRN,\
R_{\myv{m}}(\anga)R_{\myv{l}}(\angb)R_{\myv{m}}(\angc) = R_{\myv{n}}(\theta)R_{\myv{m}}(\phi)
\end{equation}
implies $f(\anga,\angb,\delta) \ge \delta$.
This can be shown easily with Corollary~\ref{coro:8}, (ii).
\enproof

\mynoindent%
{\em Proof 2.}\/
We shall show that (\ref{eq:case2_1}), i.e.,
\begin{equation}\label{eq:case2_1a}
\exists \theta,\tilde{\angc}\in\SRN,\
R_{\myv{m}}(\anga)R_{\myv{l}}(\angb)R_{\myv{m}}(\tilde{\angc}) = R_{\myv{n}}(\theta),
\end{equation}
implies 
$f(\anga,\angb,\delta) = \delta$, which is enough.
Note that $f(\anga,\angb,\delta)=\angb'$ for the angle $\angb'\in [0,\pi]$ such that
\begin{equation}\label{eq:const_evenExist}
\exists \anga',\angc'\in\SRN,\
R_{\myv{l}}(-\delta) R_{\myv{m}}(\anga) R_{\myv{l}}(\angb) R_{\myv{m}}(\angc) = R_{\myv{m}}(\anga') R_{\myv{l}}(\angb') R_{\myv{m}}(\angc')
\end{equation}
(Proof of Lemma~\ref{lem:1} in Section~\ref{sssub:ProofTh2}).
From (\ref{eq:case2_1a}) and (\ref{eq:const_evenExist}), we have
\[
\exists \anga',\angc',\tilde{\angc},\theta \in\SRN,\
R_{\myv{m}}(\anga') R_{\myv{l}}(\angb') R_{\myv{m}}(\angc'-\angc+\tilde{\angc}) = R_{\myv{l}}(-\delta) R_{\myv{n}}(\theta) ,
\]
which is, by Lemma~\ref{lem:transR3}, equivalent to
\begin{equation}\label{eq:app_eqccc}
\exists \anga',\angc',\tilde{\angc},\theta \in\SRN,\
R_{z}(\anga') R_{y}(\angb') R_{z}(\angc'-\angc+\tilde{\angc}) = R_{y}(-\delta) R_{\myv{v}}(\theta) 
\end{equation}
where $\myv{v}=(\sin \delta, 0, \cos \delta)\transp$. The absolute value of the $(1,1)$-entry of the right-hand side
in (\ref{eq:app_eqccc})
equals $\cos(\delta/2)$ since $R_{y}(-\delta)R_{v}(\theta)=R_{z}(\theta)R_{y}(-\delta)$,
which is equivalent to the equation $R_{v}(\theta)=R_{y}(\delta)R_{z}(\theta)R_{y}(-\delta)$ used before.
In view of (\ref{eq:uni1}),
this implies $\angb' = \delta$, i.e., $f(\anga,\angb,\delta) = \delta$ as desired. 
\enproofwosp

\sectionapp{Proof of (\ref{eq:a_even}) \label{app:tech}}

Observe that $\varM{\myv{n},\myv{m}}{\rm even}(U)
=\varM{\myv{m},\myv{n}}{\rm even}(U^{\dagger})$ for any $U \in {\rm SU}(2)$, by definition,
and also that
$(U^{\myv{m},\myv{l}}_{\anga,\angb,\angc})^{\dagger} = U^{\myv{m},\myv{l}}_{-\angc,-\angb,-\anga}
=U^{\myv{m},\myv{l}}_{-\angc-\pi,\angb,-\anga+\pi}$ for any $\anga,\angc$, and $\angb \in [0,\pi]$,
cf.\ footnote~\ref{fn:pipi}.
These facts give $\varM{\myv{n},\myv{m}}{\rm even}(U_{\anga,\angb,\angc}^{\myv{m},\myv{l}})
= g(-\angc-\pi,\angb,\delta) = g(\angc,-\angb,\delta)$ as desired.%
\footnote{%
As a check, one can show, using Corollary~\ref{coro:8}, (iii), that $\varM{\myv{n},\myv{m}}{\rm even}(U_{\anga,\angb,\angc}^{\myv{m},\myv{l}})=4$
if $f(\angc,-\angb,\delta)<\delta$ in the same way as in \refappcomma{app:case2}.}

\sectionapp{Proof that $\varM{\myv{m},\myv{n}}{\rm even}\big(F(U)\big)=\varM{\myv{m},\myv{n}}{\rm even}(U)$ and $N_{\myv{m},\myv{n}}\big(F(U)\big) = N_{\myv{m},\myv{n}}(U)$ \label{app:N_mnUandD}}

Let any $\myv{m},\myv{n}\in S^2$ with
$\abs{\rinnpr{\myv{m}}{\myv{\varvn}}} < 1$ and $U \in {\rm SU}(2)$ be given.
By definition, $\varM{\myv{m},\myv{n}}{\rm even}\big(F(U)\big) \le \varM{\myv{m},\myv{n}}{\rm even}(U)$. We shall show the inequality in the other direction
using the following lemma.
\begin{lemma}
For any $U,V\in {\rm SU}(2)$, $F(U)=F(V)$ iff 
$U=\pm V$. 
\end{lemma}

\mynoindent%
{\em Proof.}\/
This directly follows from the well-known fact that the kernel of $F$ is $\{ I, -I \}$,
which can be checked with (\ref{eq:Rtheta}).
\enproof

From this lemma, it readily follows that if there exist some $j \in \SNN$, $\myv{v}_1,\dots,\myv{v}_j\in S^2$, and $\phi_1, \dots, \phi_j$ $\in \SRN$ such that 
$F(U)= F\big(R_{\myv{v}_1}(\phi_1)\big) \cdots F\big(R_{\myv{v}_{j}}(\phi_j)\big)$, 
then
$U= \pm R_{\myv{v}_1}(\phi_1) \cdots R_{\myv{v}_{j}}(\phi_j)$.
But 
$ 
 -R_{\myv{v}_1}(\phi_1) \cdots R_{\myv{v}_{j}}(\phi_j)
=R_{\myv{v}_1}(\phi_1+2\pi) R_{\myv{v}_2}(\phi_2)$ $\cdots R_{\myv{v}_{j}}(\phi_j)$. 
This implies $\varM{\myv{m},\myv{n}}{\rm even}\big(F(U)\big) \ge \varM{\myv{m},\myv{n}}{\rm even}(U)$,
and hence, $\varM{\myv{m},\myv{n}}{\rm even}\big(F(U)\big) = \varM{\myv{m},\myv{n}}{\rm even}(U)$.
We also have $N_{\myv{m},\myv{n}}\big(F(U)\big) = N_{\myv{m},\myv{n}}(U)$, etc., similarly.

\section{Proof of Theorem~\ref{th:Decomp} \label{ss:proofthDecomp}}


Put
\[ \delta=\arccos\abs{\rinnpr{\myv{m}}{\myv{\varvn}}} \in  (0,\pi/2]. \]
Note
$N_{-\myv{m},\myv{\varvn}}(U) = N_{\myv{m},\myv{n}}(U)$ by definition.
Hence, we shall prove the statement assuming $\rinnpr{\myv{m}}{\myv{n}} \ge 0$, which is enough.

First, we give another corollary to Lemma~\ref{lem:1}. 
\begin{corollary}\label{coro:2}
For any $\anga,\angc\in\SRN$, and for any $\beta\in [0,\pi]$,
\[
N_{\myv{m},\myv{n}} (U_{\anga,\angb,\angc}^{\myv{m},\myv{l}}) \le 
\varMcomb{\myv{m},\myv{n}} (U_{\anga,\angb,\angc}^{\myv{m},\myv{l}}) \le 
\min \Big\{ 2 \Big\lceil \frac{\beta}{2\delta} \Big\rceil +1 , \, 
\max_{\anga'\in\SRN} g(\anga',\angb,\delta)
 \Big\} .
\]
\end{corollary}

\mynoindent%
{\em Proof.}\/
The first inequality follows from the definitions of $N_{\myv{m},\myv{n}}$ and $\varMcomb{\myv{m},\myv{n}}$.
The second inequality immediately follows from Corollary~\ref{coro:1}.
\enproof

It is easy to show, using Corollary~\ref{coro:2}, that 
\begin{equation}\label{eq:ineq_th}
N_{\myv{m},\myv{n}} \big(F(U)\big) \le N_{\myv{m},\myv{n}} (U) \le 
\nu+1 
\end{equation}
for any $U\in{\rm SU}(2)$,
where $ 
\nu := \lceil \pi/\delta \rceil $.
But 
we have $\nu+1 \le N_{\myv{m},\myv{n}}\big( F(U)\big)$ and therefore,
the equality among all
sides of (\ref{eq:ineq_th}) for
\[
U 
=\begin{cases}
R_{\myv{m}}(\pi)R_{\myv{l}}(\pi -\delta) & \mbox{if $\nu$ is even}\\
R_{\myv{l}}(\pi)  & \mbox{if $\nu$ is odd.}
\end{cases}
\]
Thus, we have proved Theorem~\ref{th:Decomp} elementarily.

\end{document}